\documentclass[aps,prl,twocolumn,10pt,bibnotes,footinbib]{revtex4-1} 
\usepackage{graphicx}
\usepackage{amsmath,ulem}
\usepackage{color}
\usepackage{lineno}
\usepackage{epstopdf}
\usepackage{units}

\def\41K{$^{41}$K}
\def\39K{$^{39}$K}
\def\87Rb{$^{87}$Rb}

\def\EE#1{\times 10^{#1}}
\def\ket#1{\left|#1\right\rangle}

\begin{document}
\title{Universal three-body physics in ultracold KRb mixtures}

\author{L. J. Wacker$^1$, N. B. J{\o}rgensen$^1$, D. Birkmose$^1$, N. Winter$^1$, M. Mikkelsen$^{2}$, J. Sherson$^1$, N. Zinner$^1$ and J. J. Arlt$^1$}

\affiliation{$^1$ Institut for Fysik og Astronomi, Aarhus Universitet, Ny Munkegade 120, DK-8000~Aarhus C, Denmark}
\affiliation{$^2$ Okinawa Institute of Science and Technology, Onna, Okinawa 904-0495, Japan}

\date{\today}

\begin{abstract}
Ultracold atomic gases have recently become a driving force in few-body physics due to the observation of the Efimov effect. While initially observed in equal mass systems, one expects even richer few-body physics in the heteronuclear case. In previous experiments with ultracold mixtures of potassium and rubidium, an unexpected non-universal behavior of Efimov resonances was observed. In contrast, we measure the scattering length dependent three-body recombination coefficient in ultracold heteronuclear mixtures of $^{39}\mathrm{K}$-\87Rb and $^{41}\mathrm{K}$-\87Rb and do not observe any signatures of Efimov resonances for accessible scattering lengths in either mixture. Our results show good agreement with our theoretical model for the scattering dependent three-body recombination coefficient and reestablish universality across isotopic mixtures. 
\end{abstract}

\maketitle

The richness of the quantum mechanical few-body problem lies in its mix of conceptual simplicity and surprising ensuing complexity. Ultracold atoms have recently become a preferred tool for the investigation of such few-body systems due to the precise available control over the two-body interaction strength. The prime example of current scientific interest is the Efimov effect~\cite{Efimov1970}. In this scenario, a short-range two-body interaction leads to an intricate spectrum of three-body bound states described by universal scaling factors, while all the two-body subsystems are unbound. Studying the consequences of this effect on corresponding many-body systems is an important yet challenging task~\cite{Zinner2014,Piatecki2014,Levinsen2015}. 

The effect was initially proposed by V. Efimov in the context of nuclear physics~\cite{Efimov1970}, however the detection in nuclei has failed so far~\cite{Jensen2004}. The first unambiguous detection of the Efimov effect was made by observing atom loss resonances in a cold cesium gas~\cite{Kraemer2006}. This led to a number of observations of this effect in single species experiments~\cite{Kraemer2006,Huckans2009,Ottenstein2008,Zaccanti2009,Pollack2009,Gross2009,Wenz2009}, in collisions between atoms and dimers~\cite{Knoop2009,Lompe2010,Nakajima2010}, the direct association of trimers~\cite{Lompe2010as,Nakajima2011,Matchey2012} and the observation of universality across different Feshbach resonances and atomic species~\cite{Gross2010,Berninger2011,Roy2013,Wang2014}. 
Recently, experiments in cesium~\cite{Huang2014} extended the range to a second loss resonance, confirming the universal scaling factor of $22.7$ between the involved Efimov trimer states~\cite{Braaten2006}. Moreover beams of molecular helium~\cite{Kunitski2015} allowed an observation of the spatial size and structure of the trimers.

Mass-imbalanced systems can provide a much denser spectrum when all scattering lenghs are large~\cite{Mikkelsen2015,Petrov2015}. So far only four different heteronuclear isotopic mixtures have been investigated. Initial experiments in the \41K-\87Rb Bose-Bose mixture reported Rb-Rb-K and a K-K-Rb resonances and a possible signature of a K-Rb atom-dimer resonance~\cite{Barontini2009, BarontiniErratum2010}. This result was controversial, since the observation of the K-K-Rb resonance was not expected at the available sample temperature~\cite{DIncao2004}. It prompted further experiments in the $^{40}$K-\87Rb Bose-Fermi mixture~\cite{Bloom2013,Hu2014}, which surprisingly only showed an Efimov type resonance for collisions between atoms and dimers but not for atomic three-body recombination, despite the fact that similar results were expected~\cite{Wang2012}. Moreover universal scaling of this resonance position is incompatible with the resonance positions observed in \41K-\87Rb~\cite{Barontini2009, BarontiniErratum2010,Helfrich2010}. More recently, an Efimov resonance in $^7$Li-$^{87}$Rb mixtures was observed~\cite{Maier2015} and the extreme mass imbalanced case of $^6$Li-$^{133}$Cs allowed for the clear observation of multiple Efimov resonances~\cite{Pires2014,Tung2014} due to the favorable scaling factor of $4.9$~\cite{Braaten2006}. 

In heteronuclear systems a natural approach is to study different isotopic mixtures e.g. the light atom can be exchanged while retaining the scattering properties of the two heavy ones. In this sense, the case of Efimov physics in K-Rb mixtures remained inconclusive, prompting our investigation in \39K-\87Rb and \41K-\87Rb Bose-Bose mixtures, despite the fact that the disadvantageous scaling factor of $\sim 115$ and 130 respectively excludes the observation of multiple Efimov resonances at currently available sample temperatures~\cite{Braaten2006}.

These experiments with ultracold atoms are enabled by the use of atomic Feshbach resonances~\cite{Chin2010}, which allow for tuning of the scattering length $a$ to large values where Efimov three-body states are expected~\cite{Fedichev1996,Burke1999,Braaten2006}. In the vicinity of such a Feshbach resonance, the three-body recombination coefficient $\alpha_{}$ typically depends on $a^4$~\cite{Nielsen1999,Burke1999,Hammer2001,Braaten2006}. This ceases to hold very close to a resonance for large $|a|$ where the finite temperature of experiments implies a constant recombination coefficient~\cite{DIncao2004}. Likewise, the $a^4$ dependence no longer holds in the limit $|a|\sim r_0$, where $r_0$ is the range of the two-body potential, since the scattering properties then depend sensitively on the short-range details. Fortunately, Efimov three-body states can typically be studied in a window between these two limits. If such an Efimov state is formed, it decays into a deeply bound dimer and a free atom, both of which leave the system. Thus the Efimov effect leads to resonances in the three-body recombination coefficient. 

In this letter, we present our measurements of the three-body recombination coefficient in heteronuclear mixtures of $^{39}\mathrm{K}$-\87Rb and $^{41}\mathrm{K}$-\87Rb in the vicinity of three Feshbach resonances. The results show good agreement with our theoretical model of the recombination coefficient for negative scattering lengths. We do not observe any signs of Efimov resonances, confirming the expected behavior of Efimov physics across similar isotopic mixtures of cold atoms~\cite{Wang2012}. 

\begin{figure}[ht]
		\centering
	  \includegraphics[width=8.6cm]{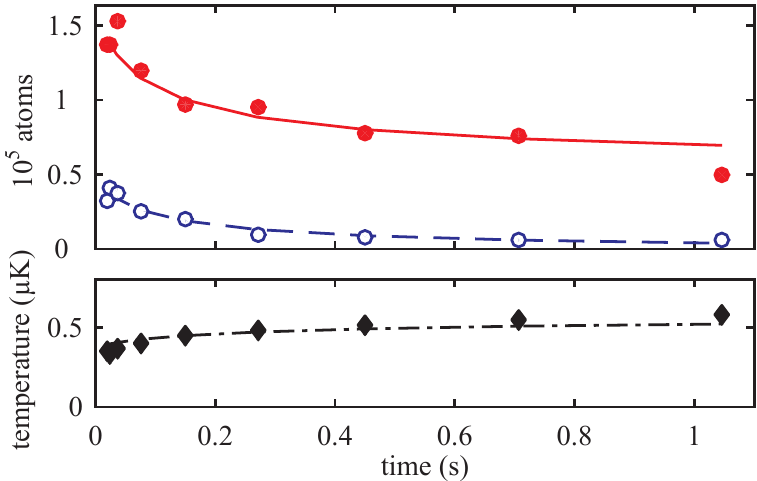}
	\caption{Atom number and temperature for a measurement of the three-body recombination coefficient. Top panel: Number of \87Rb~(solid red circles) and $^{39}\mathrm{K}$~(open blue circles) atoms. Bottom panel: Weighted average of the two cloud temperatures~(black diamonds). The lines correspond to the fitted numerical solutions of the differential equations (see text).}
	\label{fig:decay_fig}
\end{figure}

We first measured the three-body recombination coefficient in cold atomic mixtures of \39K and \87Rb. Since these experiments were in disagreement with previous results in \41K and \87Rb, we subsequently investigated these mixtures. The experiments were conducted in an apparatus previously described in~\cite{KleineBuning2010, Wacker2015}. Typically, mixtures of $0.5\EE5$~K and $1.5\EE5$~Rb atoms are prepared in a crossed beam optical dipole trap at a temperature of approximately \unit[350]{nK} in the vicinity of a Feshbach resonance. To initiate a measurement the magnetic field is abruptly changed to obtain the desired scattering length and the ensemble is held at this field for a variable duration. Subsequently the number of atoms in each species and the temperatures of the K and Rb clouds are obtained from time-of-flight images recorded after \unit[14]{ms} and \unit[16]{ms} of free expansion respectively. The hold times at each scattering length are chosen to match the timescale of the decay process. Figure \ref{fig:decay_fig} displays the typical evolution of atom numbers and temperature for \39K and \87Rb as the sample is held at a field corresponding to $a\approx\unit[-880]{{a}_0}$, where $\unit[]{a_0}$ is the Bohr radius.

The observed loss of atoms can be modeled by the coupled differential equations
\begin{align}
\label{eq:dNdt}
\frac{\mathrm{d}N_{a}}{\mathrm{d}t} =  &- \frac{2}{3} \alpha_{aab}   \int n_{a}^2 n_{b}  \mathrm{d}^3 r
	- \frac{1}{3} \alpha_{abb} \int n_{a} n_{b}^2  \mathrm{d}^3 r \nonumber \\
	&- \alpha_{aaa} \int n_{a}^3 \mathrm{d}^3 r
	- \frac{N_a}{\tau} ,
\end{align}
where $a$ and $b$ refer to the two species, $n$ are the atomic densities, $\alpha$ are the three-body recombination coefficients for the different loss channels, $\tau$ is the background lifetime and $N$ is the number of atoms.

In addition to the atom loss, two heating effects take place. Since the three-body recombination losses occur mainly at high densities, atoms at the trap center are lost preferentially, leading to heating~\cite{Weber2003}. Moreover the decay energy of the three-particle state can be released into the sample. Similar to previous work~\cite{Zaccanti2009}, the latter heating mechanism has little effect, and we neglect it in our evaluation. 

In case of $a$-$a$-$b$ recombination in mixtures, the mean potential energy of lost atoms is $\beta_{aab} = \frac{2}{3} \beta_{a} + \frac{1}{3} \beta_{b}$  where $\beta_{a} = \int n_{a}^2 n_{b} U_{a} \mathrm{d}^3 r / \int n_{a}^2  n_{b} \mathrm{d}^3 r $ and $\beta_{b} = \int n_{a} n_{b}^2 U_{b} \mathrm{d}^3 r / \int n_{a}  n_{b}^2 \mathrm{d}^3 r $, where $U$ is the potential energy, resulting in a excess energy of $\frac{3}{2} k_\text{B} T - \beta_{aab}$. When compared to the average energy $3k_\text{B} T$ of the sample of $N_a+N_b$ atoms, the heating from the $a$-$a$-$b$ loss channel is
\begin{align}
\label{eq:dTdt}
\left[ \frac{\mathrm{d}T}{\mathrm{d}t} \right]_{aab} = \alpha_{aab} \frac{\frac{3}{2} k_\text{B} T - \beta_{aab}}{3 k_\text{B}} \frac{\int n_{a}^2 n_{b}  \mathrm{d}^3 r}{N_a+N_b}, 
\end{align}
and the total heating is the sum of the heating from all channels~\cite{Weber2003,Zaccanti2009}.

An important aspect in the analysis of three-body losses in Bose-Bose systems, is the relative strength of the two interspecies three-body recombination coefficients $\alpha_{aab}$ and $\alpha_{abb}$. Since two heavy and one light atom are more strongly bound than one heavy and two light atoms, the K-Rb-Rb channel is expected to be significantly stronger~\cite{Braaten2006,Wang2012,Mikkelsen2015}. Moreover we performed our experiments with at least twice the number of Rb with respect to K atoms. Hence the losses from the K-K-Rb channel are insignificant, and are neglected in the evaluation. The single species three-body recombination coefficients are known from previous measurements~\cite{Marte2002,Zaccanti2009,Kishimoto2009} and the background lifetime was measured independently~\footnote{The single species three-body recombination coefficients for the respective scattering lengths are $\alpha_{RbRbRb} = \unit[3.2 \cdot 10^{-29}]{\frac{cm^6}{s}}$ for $^{87}\mathrm{Rb}$~\cite{Marte2002}, $\alpha_{KKK} = \unit[1 \cdot 10^{-29}]{\frac{cm^6}{s}}$ for $^{39}\mathrm{K}$~\cite{Zaccanti2009} and $\alpha_{KKK} = \unit[4 \cdot 10^{-29}]{\frac{cm^6}{s}}$ for $^{41}\mathrm{K}$~\cite{Kishimoto2009}. Note that these coefficients are at least one order of magnitude smaller than $\alpha_{KRbRb}$. Independent measurements yield lifetimes $\tau \approx \unit[100]{sec}$ for both species in our apparatus. }. This allows us to extract the three-body recombination coefficient $\alpha_{KRbRb}$ by fitting the resulting three coupled differential equations to our experimental data as shown in Fig. \ref{fig:decay_fig}.

\begin{figure*}[t!]
		\centering
	  \includegraphics[width=17.8cm]{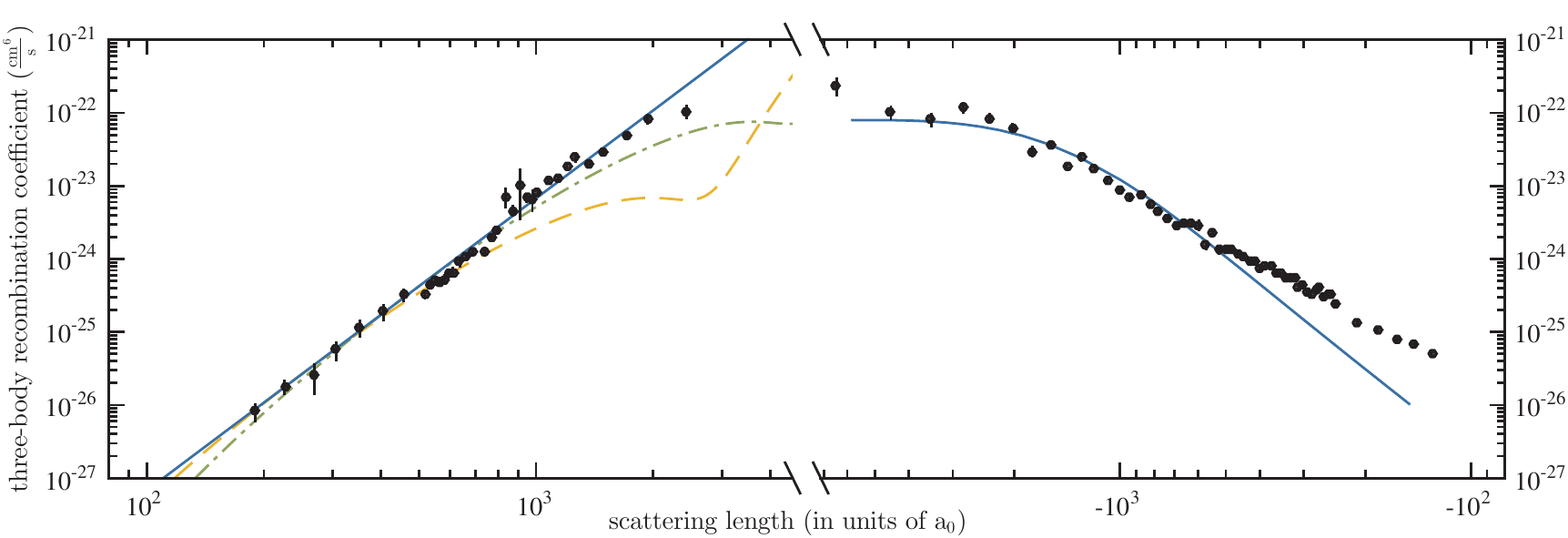}
	\caption{Three-body recombination coefficient as a function of scattering length for the \39K-\87Rb mixture. The solid blue line at negative scattering lengths is obtained from our theoretical model. For positive scattering lengths a fit with an $a^4$ dependence (blue line), and possible Efimov recombination minima at $\unit[2800]{\mathrm{a}_0}$ (dashed yellow line) and $\unit[5100]{\mathrm{a}_0}$ (dashed-dotted green line) are displayed~(see text). The error bars represent weighted averages of the fit uncertainties.}
	\label{fig:K39loss}
\end{figure*}

In our first set of experiments \39K and \87Rb atoms are both prepared in the $\ket{F,m_F} = \ket{1,-1}$ state and an s-wave Feshbach resonance at \unit[117.56]{G} is used to control the interaction~\footnote{The position of the Feshbach resonance was determined by RF molecule association spectroscopy~\cite{Klempt2008}. The scattering length for positive values was directly determined from the binding energy $E_b = \frac{\hbar^2}{2\mu a^2}$~\cite{Chin2010}, while for negative and low positive values the functional dependence $a(B)= a_{bg} \left[ 1-(B_c-B_0)/(B-B_c) \right]$ with $B_0=\unit[116.4]{G}$ and $B_c=\unit[117.5]{G}$ was used~\cite{Simoni2008, Wacker2015}.}. The sample is held in an optical dipole potential with trapping frequencies $\nu_\rho= \unit[118]{Hz}~(\unit[84]{Hz})$ and $\nu _z= \unit[164]{Hz}~(\unit[111]{Hz})$ for \39K~(\87Rb). Decay measurements are performed at various interaction strengths and by using the fitting procedure outlined above, the resulting $\alpha_\text{KRbRb}$ is obtained as shown in Fig.~\ref{fig:K39loss}. Besides the expected increase in the recombination coefficient for positive and negative scattering lengths, no significant enhanced losses due to an Efimov resonance or suppressed losses due to an Efimov recombination minimum are observed~\cite{Nielsen1999}.

\begin{figure*}[htbp]
		\centering
	  \includegraphics[width=17.8cm]{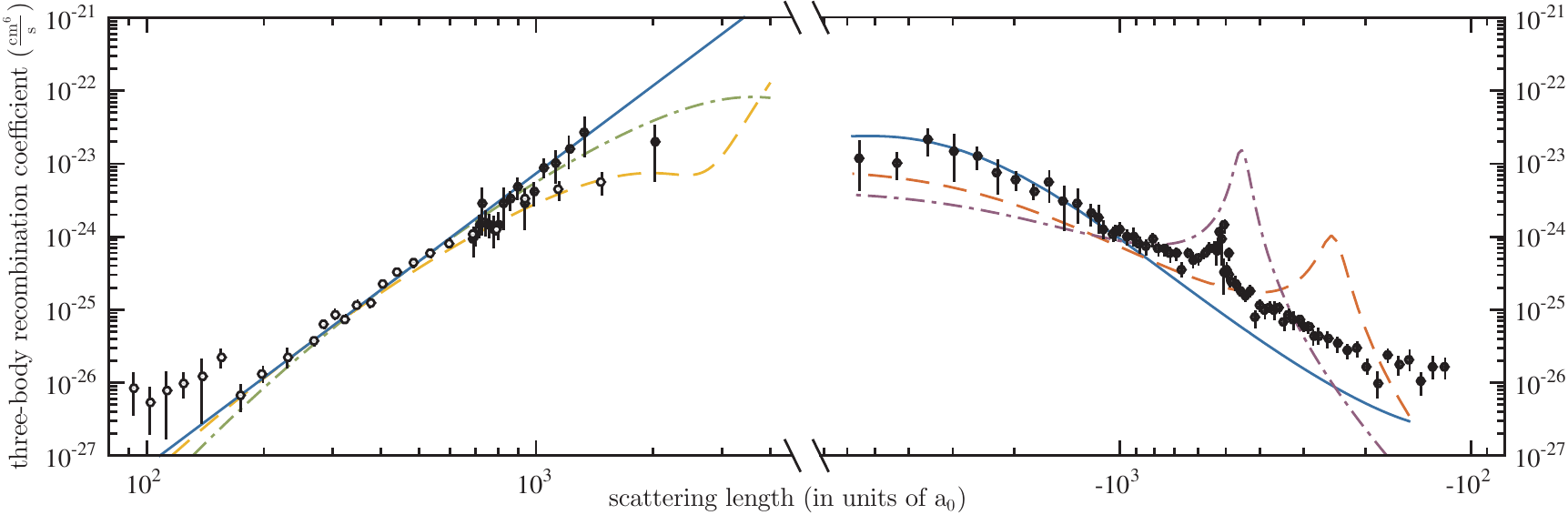}
	\caption[dlabel]{Three-body recombination coefficient as a function of scattering length for the \41K-\87Rb mixture recorded using two Feshbach resonances at $\unit[38]{G}$ (solid circles) and $\unit[79]{G}$ (open circles). At negative scattering lengths, our theoretical model is used to display three different scenarios: no Efimov resonance (solid blue line), a resonance at $\unit[-246]{\mathrm{a}_0}$ (dashed red line) and a resonance at $\unit[-450]{\mathrm{a}_0}$ (dashed-dotted purple line). For positive scattering lengths a fit with an $a^4$ dependence (blue line), and possible Efimov recombination minima at $\unit[2800]{\mathrm{a}_0}$ (dashed yellow line) and $\unit[5100]{\mathrm{a}_0}$ (dashed-dotted green line) are displayed~(see text). The error bars represent weighted averages of the fit uncertainties.}
	\label{fig:K41loss}
\end{figure*}

In our second set of experiments, \41K and \87Rb atoms are prepared in the $\ket{1,1}$ state in an optical dipole trap with trapping frequencies of $\nu_\rho= \unit[124]{Hz}~(\unit[89]{Hz})$ and $\nu _z= \unit[170]{Hz}~(\unit[119]{Hz})$ for \41K~(\87Rb). We primarily employ a Feshbach resonance at $\unit[38]{G}$ to tune the scattering length. Since scattering lengths between $\unit[0]{}$ and $\unit[640]{\mathrm{a}_0}$ can not be addressed with this resonance due to a large offset at low fields, we additionally use a resonance at $\unit[79]{G}$ to address scattering lengths in this range~\cite{Thalhammer2009}. The scattering length dependence on the magnetic field is obtained from a model for the full molecular potential~\cite{Tiemann2015}. The three-body recombination coefficients obtained by the fitting procedure outlined above are shown in Fig.~\ref{fig:K41loss}. Once again we do not observe any significant deviations from the expected $a^4$ dependence for positive scattering lengths and thus no sign of Efimov resonances.

\begin{figure}[ht]
		\centering
	  \includegraphics[width=8.6cm]{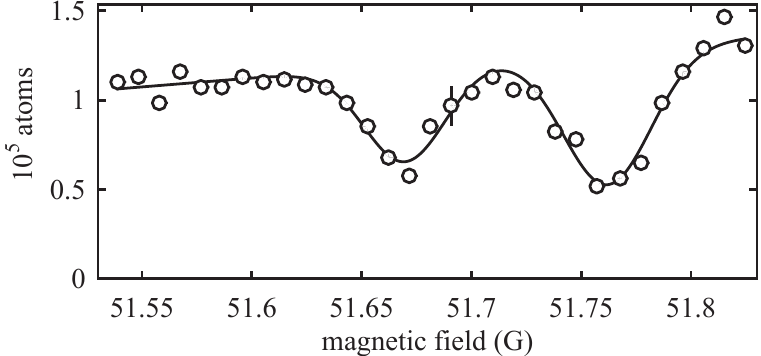}
	\caption{Collisional spectroscopy of a p-wave Feshbach resonance. Summed atom number of $^{41}\mathrm{K}$ and \87Rb remaining in the trap for a fixed hold time as a function of magnetic field. The solid line is a fit of two Gaussians on a linear background. A typical error bar is displayed.}
	\label{fig:p_wave}
\end{figure}

However, we do observe a loss peak at approximately $\unit[-500]{\mathrm{a}_0}$. We have performed collisional spectroscopy at this feature and clearly resolve the double peak structure shown in Fig.~\ref{fig:p_wave}. This identifies this feature as a p-wave Feshbach resonance in correspondence with theoretical prediction~\cite{Thalhammer2009}. Besides this Feshbach resonance, we do not observe any significant enhancement of losses due to the Efimov scenario for negative scattering lengths. 

For both experiments the average loss ratio between the two species is observed to be $\Delta N_{Rb}/\Delta N_{K} \approx 2$, confirming that the recombination of Rb-Rb-K is the main loss channel. The lower limit for the detection of a resonance in both scenarios is approximately $|a| \approx 2R_{vdW}\approx \unit[144]{\mathrm{a}_0}$ where the details of the interatomic potential become dominant. The upper limit is set by the temperature at approximately $\unit[2000]{\mathrm{a}_0}$ where the recombination coefficient saturates~\cite{DIncao2004}. The deviation of our data from the expected behavior is close to these limits. However, there are still some discrepancies due to the interaction between the rubidium atoms in the three-body recombination process which have not been included in the theoretical model. We estimate a 15\% systematic uncertainty of the atom number. Equation \eqref{eq:dNdt} thus yields a relative uncertainty of the three-body recombination coefficient of 50\%. This corresponds to a small overall shift of the data and does not influence our conclusions.

In order to exclude the influence of further Feshbach resonances we have performed collisional spectroscopy at the predicted resonance positions~\cite{Thalhammer2009} in the range of our experiment. We identified three previously unobserved resonance positions (see Supplemental Material~\cite{Wacker2016SupMat}), which are in excellent agreement with prediction. These resonances are in general much narrower than the employed s-wave resonances and thus only influence the three-body recombination rate in a very small region close to their center~\cite{Chin2010}.

In the following we compare our three-body recombination results with our theoretical model and with results obtained in $^{41}\mathrm{K}$-\87Rb and $^{40}\mathrm{K}$-\87Rb~\cite{Barontini2009, BarontiniErratum2010,Bloom2013,Kato2015}. For negative scattering lengths, our theoretical framework~\cite{Mikkelsen2015} is applied, assuming no Efimov resonances as shown in Fig.~\ref{fig:K39loss} and \ref{fig:K41loss}. The theory is based on an optical model approach with an imaginary potential at short distances and fully accounts for finite temperature effects by considering recombination at all energies and folding the results with a Boltzmann distribution for the experimentally determined temperature. Our model does not involve any rescaling and gives the absolute value of the recombination rate as a function of the input parameters of the optical model. Our data agrees well with the theoretical prediction for a broad range of scattering lengths. For low scattering lengths the model is less accurate since it does not take the details of the interatomic potential at short range into account. We observe good agreement with an $a^4$ dependence for positive scattering lengths for intermediate scattering lengths, while it deviates for large scattering lengths due to thermal saturation and for small scattering lengths due to the short range details of the interatomic potential.

All K-\87Rb combinations have comparable van-der Waals interaction characterized by $R_{vdW}\approx$~\unit[72]{$\mathrm{a}_0$}~\cite{Chin2010,Simoni2008}. In previous experiments with \41K-\87Rb mixtures~\cite{Barontini2009, BarontiniErratum2010}, loss maxima were found at $a_-=\unit[-246]{\mathrm{a}_0}$ for the K-Rb-Rb and at $a_-=\unit[-22000]{\mathrm{a}_0}$ for the K-K-Rb resonance. The first of these resonances is absent in our results and the saturation of the tree-body recombination rate confirms that the K-K-Rb resonance can not be observed at currently available sample temperature~\cite{DIncao2004}. 

In the following, we interpret our results in view of universal scaling. Based on collisions between atoms and dimers, atom-dimer resonances were observed at $a_*=\unit[230]{\mathrm{a}_0}$ in $^{40}\mathrm{K}$-\87Rb mixtures~\cite{Bloom2013} and more recently at $a_*=\unit[360]{\mathrm{a}_0}$ in $^{41}\mathrm{K}$-\87Rb mixtures~\cite{Kato2015}. If these atom-dimer resonances are associated with the lowest Efimov state, universal scaling allows for an estimate of  atomic Efimov resonance positions at $a_-=\unit[-55000]{\mathrm{a}_0}$ and $a_-=\unit[-91000]{\mathrm{a}_0}$ respectively. Thus these resonances lie at large interaction strengths outside the currently accessible regime~\cite{Helfrich2010}. This scenario is fully compatible with our results.

In a second scenario the atom-dimer resonances could be connected with the first excited Efimov state. In this case an estimate based on universal scaling suggests atomic Efimov resonances at $a_-=\unit[-450]{\mathrm{a}_0}$~\cite{Bloom2013} and $a_-=\unit[-690]{\mathrm{a}_0}$~\cite{Helfrich2010}. Figure~\ref{fig:K41loss} shows our theoretical model with added resonances at $\unit[-246]{\mathrm{a}_0}$ and $\unit[-450]{\mathrm{a}_0}$. Even within the uncertainty of universal scaling this is in disagreement with our data and shows that this scenario is unlikely~\cite{Zenesini2014}.

For positive scattering lengths the observation of recombination minima is possible, depending on the position of the atomic Efimov resonances. Our data for \39K-\87Rb and \41K-\87Rb does not show such minima, supporting the scenario that the atomic Efimov resonances lie at large interaction strengths. In this case possible positions for the minima at $a_+=\unit[2800]{\mathrm{a}_0}$~\cite{Wang2012} and $a_+=\unit[5100]{\mathrm{a}_0}$~\cite{Bloom2013} have been suggested. We show these minima~\footnote{We use the analytical expressions given in~\cite{Helfrich2010} with $\eta^*= 0.02$~\cite{Bloom2013}. The coefficient $C_\alpha$ is obtained from a fit to our data with $C_\alpha = 529$ for the \39K-\87Rb and $C_\alpha = 575$ for the \41K-\87Rb mixture.} in Fig.~\ref{fig:K39loss} and~\ref{fig:K41loss} using the analytical expressions from~\cite{Helfrich2010}. In \39K-\87Rb mixtures our data indicates that recombination minima lie beyond $\unit[5000]{\mathrm{a}_0}$, consistent with the scenario that atomic Efimov resonances lie at large interaction strengths~\cite{Ulmanis2016}.

In conclusion, we have measured the three-body recombination coefficient over a range of four orders of magnitude for positive and negative interaction in mixtures of $^{39}\mathrm{K}$-\87Rb and $^{41}\mathrm{K}$-\87Rb. The observed behavior for negative scattering lengths is in good agreement with our theoretical model and does not reproduce the results of reference~\cite{Barontini2009, BarontiniErratum2010}. Both for negative and positive scattering lengths, no distinct features, which could be associated with the presence of an Efimov resonance or a recombination minimum, were observed in the range of $\unit[144]{\mathrm{a}_0}$ to $\unit[2000]{\mathrm{a}_0}$. These results resolve a debate in three-body physics of ultracold gas mixtures, and hence contribute to the general understanding of mass-imbalanced few-body systems.

\begin{acknowledgments}
We thank E. Tiemann for a prediction of the magnetic field dependent scattering lengths, fruitful discussions with F. Minardi and C. Greene for comments on the manuscript. We thank the Danish Council for Independent Research, and the Lundbeck Foundation for support.
\end{acknowledgments}

\bibliography{MIXEfimov}

\section{Supplementary Material}
\section{Observed Feshbach resonances}

\begin{table}[htbp]
	\centering
	\begin{tabular}{ c   c   c }
		\hline\hline
		$\text{assignment}$ & $B_\text{th}$ (\unit{G}) & $B_\text{exp}$ (\unit{G}) \\ \hline
		(331)/(202) & 48  & 47.95 \\ 
		(321) & 52  & 51.67(51.76) \\ 
		(112) & 65  & 64.73 \\ 
		(102) & 73  & 72.74 \\ \hline\hline
	\end{tabular}
	\caption{Overview of the observed and predicted Feshbach resonances \41K and \87Rb. Assignments of the quantum numbers ($f m_f l'$) are from \cite{Thalhammer2009}. The resonance at 48~\unit{G} has been observed previously~\cite{Thalhammer2009}.}
	\label{tab:higherorderfeshbachresonances}
\end{table}

\end{document}